\documentclass[12pt]{article}
\usepackage{amssymb,amsmath,cite,epsfig,graphicx} 
\catcode`\@=11
\@addtoreset{equation}{section}
\def\theequation{\arabic{section}.\arabic{equation}}
     
\catcode`\@=11
\def\thesection{\arabic{section}.}

\def\appendix{\setcounter{section}{0}
        \def\thesection{Appendix.}
        \def\theequation{\Alph{section}.\arabic{equation}}}
\def\section{\@startsection{section}{1}{\z@}{3.5ex plus 1ex minus
   .2ex}{2.3ex plus .2ex}{\large\bf}}

\newcommand{\beq}{\begin{equation}}
\newcommand{\eeq}{\end{equation}}

\begin{document}
\thispagestyle{empty}
\vspace{.5in}
\begin{flushright}
UCD-04-23\\
August 2004\\
revised January 2005\\
hep-th/0408123\\
\end{flushright}
\vspace{.5in}
\begin{center}
{\Large\bf
  Horizon Constraints and Black Hole Entropy}
\end{center} 
\vspace*{3ex}
\begin{center}
{S.~C{\sc arlip}\footnote{\it email: carlip@physics.ucdavis.edu}\\
       {\small\it Department of Physics}\\
       {\small\it University of California}\\
       {\small\it Davis, CA 95616}\\{\small\it USA}}
\end{center} 
\vspace*{3ex}
\begin{center}
{\large\bf Abstract}
\end{center}
\begin{center}
\begin{minipage}{5in}
{\small A question about a black hole in quantum gravity is a
conditional question: to obtain an answer, one must restrict initial 
or boundary data to ensure that a black hole is actually present.  
For two-dimensional dilaton gravity---and probably for a much wider 
class of theories---I show that the imposition of a spacelike 
``stretched horizon'' constraint modifies the algebra of symmetries, 
inducing a central term.  Standard conformal field theory techniques 
then fix the asymptotic density of states, successfully reproducing 
the Bekenstein-Hawking entropy.  The states responsible for black hole 
entropy can thus be viewed as ``would-be gauge'' states that become 
physical because the symmetries are altered.
}
\end{minipage}
\end{center}
\vfill
\addtocounter{footnote}{-1}
\addtocounter{page}{-1}
 
\newpage

Suppose one wishes to ask a question about a quantum black hole.  In a 
semiclassical theory, this is straightforward, at least in principle---one 
can look at quantum fields and gravitational perturbations around a black 
hole background.  In a full quantum theory of gravity, though, such a 
procedure is no longer possible---there is no fixed background, and the 
theory contains both states with black holes and states with none.  One 
must therefore make one's question conditional: ``\emph{If} a black hole 
with property \emph{X} is present\dots''  Equivalently, one must impose 
constraints, either on initial data or on boundaries, that restrict the 
theory to one containing an appropriate black hole \cite{Carlip1}.

Classical general relativity is characterized by a symmetry algebra, the
algebra of diffeomorphisms.  But it is well known that the introduction 
of new constraints can alter such an algebra \cite{Dirac,Hen,Carlip5}.  
For the simple model of two-dimensional dilaton gravity, I will show 
below that the imposition of suitable ``stretched horizon'' constraints 
has the effect of adding a central extension to the algebra of 
diffeomorphisms of the horizon.  This is a strong result, because such 
a centrally extended algebra is powerful enough to almost completely fix 
the asymptotic behavior of the density of states, that is, the entropy 
\cite{Cardy,Cardy2}.  Indeed, I will show that given a reasonable 
normalization of the ``energy,'' standard conformal field theoretical 
methods reproduce the correct Bekenstein-Hawking entropy.  Moreover, 
while the restriction to two-dimensional dilaton gravity is a significant 
one, I will argue that the conclusions are likely to extend to much more 
general settings.

These results suggest that the entropy of a black hole can be explained
by two key features: the imposition of horizon boundary conditions, which 
can alter the physical content of the theory by promoting ``pure gauge''
fields to dynamical degrees of freedom, and the existence of a Virasoro
algebra, which can control the asymptotic density of states.  Both of
these features are present for the (2+1)-dimensional black hole \cite{%
Strominger,Birmingham,Carlip2}, and a number of authors---see, for example, 
\cite{Carlip3,Solodukhin,Cadoni,Navarro,Birm,Gupta,Giacomini,Frolov,Cvitan,%
Camblong}---have suggested that near-horizon symmetries may control generic 
black hole entropy.  In particular, we shall see that the near-horizon 
conformal symmetry of \cite{Carlip4} is closely related to the horizon 
constraint introduced here.

\section{Dilaton Gravity in a Null Frame}

We start with canonical two-dimensional dilaton gravity in a null 
frame, that is, expressed in terms of a null dyad $\{l^a,n^a\}$ with 
$\ell\cdot n = -1$.  The metric is then
\beq
g_{ab} = -l_an_b - l_bn_a ,
\label{a1}
\eeq
and ``surface gravities'' $\kappa$ and $\bar\kappa$ may be defined by
\begin{align}
&\nabla_al_b = -\kappa n_al_b - {\bar\kappa}l_al_b \nonumber\\
&\nabla_an_b = \kappa n_an_b + {\bar\kappa}l_an_b ,
\label{a2}
\end{align}
where the second of eqns.\ (\ref{a2}) follows from the first.  It is easy
to check---for instance, by computing $[\nabla_a,\nabla_b]l^b$---that the
scalar curvature is
\beq
R = 2\nabla_a(\kappa n^a - {\bar\kappa}l^a) 
  = -\frac{2}{\sqrt{-g}}\partial_a\left(\kappa{\hat\epsilon}^{ab}n_b +
    {\bar\kappa}{\hat\epsilon}^{ab}l_b\right)
\label{a3}
\eeq
where $\hat\epsilon^{ab}=-\hat\epsilon^{ba} $ is the Levi-Civita density, 
$\hat\epsilon^{uv}=1$.  

In addition to the metric, two-dimensional dilaton gravity contains a dilaton 
field, which I shall call $A$.  For models obtained by dimensionally reducing 
Einstein gravity, $A$ is just the transverse area.  With appropriate field
redefinitions \cite{Gegenberg}, the action becomes
\begin{align}
I &= \frac{1}{16\pi G}\int d^2x \sqrt{-g}\left[AR + V(A)\right] \label{a4} \\
  &= \frac{1}{16\pi G}\int d^2x \left[{\hat\epsilon}^{ab}
     \left(2\kappa n_b\partial_aA
     - 2{\bar\kappa} l_b\partial_aA\right) + \sqrt{-g}V\right]
\nonumber 
\end{align}
with a potential $V(A)$ that depends on the specific model.  If one now 
defines components 
\beq
l = \sigma du + \alpha dv, \qquad n = \beta du + \tau dv 
\label{a4a}
\eeq
with respect to coordinates $(u,v)$, and chooses units $16\pi G=1$, the 
Lagrangian becomes
\begin{align}
L = &\frac{1}{\sigma\tau - \alpha\beta}\biggl[
    2\left(\tau{\dot A}-\beta A'\right)\left({\dot\alpha}-\sigma'\right)
\nonumber\\
  &\ +2\left(\alpha{\dot A}-\sigma A'\right)\left({\dot\tau}-\beta'\right)
  \biggr]
  + (\sigma\tau - \alpha\beta)V(A) ,
\label{a5}
\end{align}
where a dot denotes a derivative with respect to $u$ and a prime a derivative
with respect to $v$.  One can immediately read off the canonical momenta,
\begin{align}
&\pi_\alpha 
  = \frac{2}{\sigma\tau - \alpha\beta}\left(\tau{\dot A}-\beta A'\right),
\nonumber\\
&\pi_\tau 
  = \frac{2}{\sigma\tau - \alpha\beta}\left(\alpha{\dot A}-\sigma A'\right),
\label{a6}\\
&\pi_A 
  = \frac{2}{\sigma\tau - \alpha\beta}\left[\tau\left({\dot\alpha}-\sigma'\right)
  + \alpha\left({\dot\tau}-\beta'\right)\right] .\nonumber
\end{align}
The variables $\sigma$ and $\beta$ appear with no time derivatives, and act 
as Lagrange multipliers.  As in any diffeomorphism-invariant theory, the 
Hamiltonian is a linear combination of constraints: $H = \sigma C_\perp + 
\frac{\beta}{\tau}\left(C_\parallel - \alpha C_\perp\right)$, with
\begin{align}
C_\perp &= \pi_\alpha{}' - \frac{1}{2}\pi_\alpha\pi_A - \tau V(A) \nonumber\\
C_\parallel &= \pi_A A' - \alpha\pi_\alpha{}' - \tau\pi_\tau{}' .
\label{a8}
\end{align}
An additional constraint appears because both $\pi_\alpha$ and $\pi_\tau$ depend 
only on $\dot A$:
\beq
C_\pi = \tau\pi_\tau - \alpha\pi_\alpha + 2A' .
\label{a9}
\eeq
$C_\perp$ and $C_\parallel$ are ordinary Hamiltonian and momentum constraints,
while $C_\pi$ is a disguised version of the generator of local Lorentz 
invariance, appearing because the pair $\{l,n\}$ is invariant under the boost
$l\rightarrow fl$, $n\rightarrow f^{-1}n$.  The constraints have the Poisson 
algebra
\begin{alignat}{2}
&\{C_\perp[\xi],C_\perp[\eta]\} = 0  & \qquad
&\{C_\parallel[\xi],C_\perp[\eta]\} = C_\perp[\xi\eta'] \nonumber\\
&\{C_\parallel[\xi],C_\parallel[\eta]\} = C_\parallel[\xi\eta'-\eta\xi'] &
&\{C_\perp[\xi],C_\pi[\eta]\} = C_\perp[\xi\eta] \nonumber\\
&\{C_\parallel[\xi],C_\pi[\eta]\} = C_\pi[\xi\eta']   &
&\{C_\pi[\xi],C_\pi[\eta]\} = 0  \label{a10}
\end{alignat}
where $C[\xi]$ denotes the ``smeared'' constraint $\int\! dv\, \xi C$ on a
surface of constant $u$.

\section{Horizon Constraints}

So far, this has all been standard, albeit in a slightly unusual parametrization.
Now, however, let us demand that the initial surface $u=0$ be a horizon,
which we can define in the sense of \cite{Ashtekar} as a null surface (with null 
normal $l^a$) with vanishing expansion $\vartheta$.  The usual definition of 
``expansion'' does not apply in two spacetime dimensions, but a straightforward 
generalization, 
\beq
\vartheta = l^a\nabla_aA/A,
\label{b0}
\eeq 
captures the same information:
it describes the fractional change in the transverse area, and is proportional
to the conventional expansion in models obtained by dimensional reduction.  The 
imposition of such a horizon constraint is a bit delicate, however, as I shall 
now describe.

First, for a standard canonical approach to work, the initial surface should 
be spacelike.  While a canonical analysis of gravity with a null slicing is 
possible \cite{Goldberg,Torre}, it introduces considerable complications.  We 
shall avoid these by considering a slightly distorted horizon that is ``almost 
null,'' requiring that $\alpha=\epsilon_1\ll1$ in (\ref{a4a}).  

For such a ``stretched horizon,'' one should not require that the expansion---here,
the logarithmic derivative of $A$ at the $u=0$ surface---vanish, but only that 
it be sufficiently small.  Here a second subtlety arises: the null normal $l^a$
does not have a unique normalization.  While the condition of vanishing expansion 
is independent of the normalization of $l^a$, the condition of ``small'' expansion 
is not; indeed, by simply rescaling $l^a$ by a constant, one can make $\vartheta$ 
arbitrarily large.  Fortunately, another quantity, the surface gravity $\kappa$, 
changes in the same way under constant rescalings of $l^a$.  We therefore require 
that $l^v\nabla_vA/\kappa A = \epsilon_2$, or, from (\ref{a6}), 
$A' - \frac{1}{2}\epsilon_2A\pi_A = 0$.  I show in the appendix that for
$\epsilon_2$ small and negative, these constraints yield a spacelike surface 
that closely traces the true horizon.

A final ambiguity comes from the existence of the constraint (\ref{a9}), which 
allows us to trade $A'$ in the expansion for the combination $\tau\pi_\tau - 
\alpha\pi_\alpha$.  It is not clear which expression provides the ``correct'' 
constraint.  We therefore leave the choice open, writing the horizon 
constraints in the form
\begin{align}
&K_1 = \alpha-\epsilon_1 = 0 \nonumber\\
&K_2 = A' - \frac{1}{2}\epsilon_2A_+\pi_A + \frac{a}{2}C_\pi = 0
\label{b1}
\end{align}
where $a$ is an arbitrary constant.
I have substituted the horizon value $A_+$ for $A$ in the second line; this
will avoid complicated field redefinitions later without changing the physics,
since the difference between $A$ and $A_+$ is ${\cal O}(\epsilon_2)$.

We may now make the fundamental observation that $\{K_i,K_j\}\ne0$, that is, that 
the $K_i$ are second class constraints.  Indeed, a straightforward computation 
yields
\beq
\{K_i(x),K_j(y)\}  = \Delta_{ij}(x,y)  
  = \left(\begin{array}{cc} 0 & -\frac{a}{2}\alpha\delta(x-y)\\
  \frac{a}{2}\alpha\delta(x-y) & -(1+a)\epsilon_2A_+\delta'(x-y)\end{array}\right) .
\label{b2} 
\eeq
For the Poisson algebra of our remaining observables to be consistent with these
constraints, we should replace Poisson brackets by Dirac brackets \cite{Dirac,Hen},
\beq
\{P,Q\}^* = \{P,Q\} \label{b3} 
   - \sum_{i,j}\int\,dx\,dy \{P,K_i(x)\}\Delta^{-1}_{ij}(x,y)\{K_j(y),Q\}
\eeq
so that $\{P,K_i(x)\}^*=0$.  Equivalently, this amounts to replacing every 
observable $P$ by a combination 
\beq
P^* = P + c_1K_1 + c_2K_2,
\label{b4}
\eeq 
with coefficients $c_i$ chosen so that $\{P^*,K_i(x)\}=0$ \cite{Bergmann}.
Since the $K_i$ vanish for physical configurations, $P^*$ is physically
equivalent to $P$; and since $P^*$ has a vanishing bracket with the $K_i$,
we can impose these constraints consistently without changing the meaning
of the Poisson bracket.

Now, the horizon constraints (\ref{b1}) hold only on the surface $u=0$, so it 
is not clear that Dirac brackets are needed for the Hamiltonian constraint
$C_\perp$, which evolves quantities off the initial surface.  The momentum and
Lorentz constraints $C_\parallel$ and $C_\pi$, on the other hand, clearly need to
be modified appropriately.  It is not hard to show that the ``starred'' constraints
are
\begin{align}
C_\parallel^* &= C_\parallel + \frac{4(1+a)}{a^2}\epsilon_2A_+
  \frac{K_1^{\prime\prime}}{\epsilon_1} - \frac{2}{a}K_2' \nonumber\\
C_\pi^* &= C_\pi + \frac{2}{a}\left(1+\frac{2}{a}\right)\epsilon_2A_+
  \frac{K_1^{\prime}}{\epsilon_1} - \frac{2}{a}K_2
\label{b4a}
\end{align}
and that the Dirac brackets become
\begin{align}
\{C_\parallel[\xi],C_\parallel[\eta]\}^* &= C_\parallel[\xi\eta'-\eta\xi'] 
  - \frac{2(1+a)}{a^2}\epsilon_2A_+\int\,dv(\xi'\eta'' - \eta'\xi'') \nonumber\\
\{C_\parallel[\xi],C_\pi[\eta]\}^* &= C_\pi[\xi\eta'] 
  +\frac{2}{a}\left(\frac{2}{a}+1\right)\epsilon_2A_+\int\,dv\,\xi'\eta' \nonumber \\
\{C_\pi[\xi],C_\pi[\eta]\}^* &= -\frac{2}{a^2}\epsilon_2A_+\int\,dv(\xi\eta' - \eta\xi') 
  . \label{b5} 
\end{align}
With the choice $a=-2$, the anomalous term in the transformation of $C_\pi$
vanishes, and this algebra has a simple conformal field theory interpretation
\cite{CFT}: the $C_\parallel$ generate a Virasoro algebra with central charge
\beq
\frac{c}{48\pi} = -\frac{1}{2}\epsilon_2A_+ ,
\label{b6}
\eeq
while $C_\pi$ is an ordinary primary field of weight one. (Recall that $\epsilon_2<0$,
so the central charge is positive.)

Thus far we have freely used integration by parts.  To compute the entropy 
associated with the horizon, we will also need the classical value of $C_\parallel^*$. 
At first sight this must vanish, since the constraint is zero for any classical
configuration.  As usual in general relativity, though, $C_\parallel^*$ has a 
nontrivial boundary term, which gives a nonvanishing classical contribution.

In particular, suppose that our ``stretched horizon'' has an end point at the actual
horizon, at $v=v_+$.  To obtain the boundary term at this point, one must decide what 
is held fixed at the boundary. Here we can take a hint from \cite{CarTeit}, where it 
is shown that the boundary conditions that give the correct horizon contribution to
the Euclidean path integral are those that hold fixed the variable conjugate to $A$.
Moreover, since we are limiting ourselves to configurations for which the constraints
$K_1$ and $K_2$ vanish, we can take $\delta K_1 = \delta K_2 =0$ at $v_+$.  With these 
boundary conditions, it is evident that the variation of $C_\parallel^*[\xi]$
contains a boundary term $\xi\pi_A\delta A|_{v_+}$, which must be canceled off to 
obtain well-defined Poisson brackets \cite{RegTeit}.  Hence
\beq
C_{\parallel\,\mathit{bdry}}^* = -\left.\xi\pi_AA\right|_{v=v_+} .
\label{b8}
\eeq

\section{Computing the Entropy}

We are now in a position to compute the entropy associated with the stretched horizon.
The key ingredient is the Cardy formula \cite{Cardy,Cardy2}, which states that for a 
conformal field theory with central charge $c$, the number of states with eigenvalue 
$\Delta$ of the Virasoro operator $L_0$ has the asymptotic form
\beq
\rho(\Delta) = \exp\left\{2\pi\sqrt{\frac{c_{\mathit{eff}}\Delta}{6}}\right\}
\label{c1}
\eeq
with $c_{\mathit{eff}} = c-24\Delta_0$, where $\Delta_0$ is the smallest eigenvalue
of $L_0$.  For us, $c$ is given by (\ref{b6}), while $\Delta$ is the boundary
contribution (\ref{b8}) to $C_\parallel^*[\xi_0]$, where $\xi_0$ is the generator 
of ``constant'' translations in $v$.  These ``classical'' values may be subject to 
quantum corrections, but for macroscopic black holes, these will be small.  Indeed, 
making the usual quantum substitution $\{\ ,\ \} \rightarrow \frac{1}{i\hbar}[\ ,\ ]$
and restoring the factors of $16\pi G$, the first equation in (\ref{b5}) becomes
\beq
\left[\frac{1}{16\pi\hbar G}C_\parallel[\xi],
  \frac{1}{16\pi\hbar G}C_\parallel[\eta]\right] 
  = \frac{i}{16\pi\hbar G}C_\parallel[\xi\eta'-\eta\xi']
  + \frac{i\epsilon_2A_+}{32\pi\hbar G}\int\,dv(\xi'\eta'' - \eta'\xi'') ,
\label{c1a}
\eeq
from which we can read off the values
\beq
c = -\frac{3\epsilon_2A_+}{2\hbar G}, \qquad
\Delta = -\frac{1}{16\pi\hbar G}\left.\xi_0\pi_AA\right|_{v=v_+} .
\label{c1b}
\eeq
The classical central charge will thus dominate as long as $|\epsilon_2|$ is 
large compared to the tiny quantity $A_{\hbox{\scriptsize Planck}}/A_+$.
For Planck-sized black holes, on the other hand, we might expect quantum 
corrections to become important, leading to a breakdown in this analysis.

It remains to determine the parameter $\xi_0$ in (\ref{c1a}).  To define energy 
in an asymptotically flat spacetime, one may fix the normalization of $\xi_0$ at 
infinity.  Working only on the stretched horizon, though, we do not have that 
luxury; as in the ``isolated horizons'' program \cite{Ashtekar}, we must cope 
with an uncertainty in the normalization of the horizon diffeomorphisms.  As noted 
in \cite{Carlip4}, though, there is one natural choice associated with a stretched 
horizon: 
\beq
z = e^{2\pi i A/A_+}
\label{c2}
\eeq
gives us a good intrinsic coordinate associated with such a surface.  We
can therefore parametrize diffeomorphisms by vector fields
\beq
\xi_n = \frac{A_+}{2\pi A'} z^n ,
\label{c3}
\eeq
where the prefactor has been chosen to ensure that $[\xi_m,\xi_n] = i(n-m)\xi_{m+n}$.
Of course, these fields are only defined on a stretched horizon---$A'$ is zero at 
the actual horizon---but our canonical formalism breaks down on the actual horizon 
as well.  (Note that $z=1$ at the horizon, so the diffeomorphisms (\ref{c3}) 
reduce to a single constant, albeit infinite, shift.)

Inserting (\ref{c3}) into (\ref{c1a}), we see that
\beq
\Delta = -\frac{1}{16\pi\hbar G}\left.\xi_0\pi_AA \right|_{v_+} 
  = -\frac{A_+^2}{32\pi^2\hbar G}\left.\frac{\pi_A}{A'} \right|_{v_+} 
  = -\frac{A_+}{16\pi^2\epsilon_2\hbar G}
\label{c4}
\eeq
where I have used the constraint $K_2=0$ in the last equality.  Combining (\ref{c1}), 
(\ref{c1b}), and (\ref{c4}), and assuming that $\Delta_0=0$, we obtain a microcanonical 
entropy
\beq
S = \ln\rho(\Delta) \label{c5}\\
  = 2\pi\sqrt{\frac{\epsilon_2A_+}{4\hbar G}\cdot
    \frac{A_+}{16\pi^2\epsilon_2\hbar G}} 
  = \frac{A_+}{4\hbar G} 
\eeq
which agrees precisely with the expected Bekenstein-Hawking entropy.  

We thus see that the entropy of a two-dimensional black hole can, indeed,
be determined almost uniquely from the imposition of appropriate 
``horizon constraints.''  The Cardy formula does not tell us \emph{what} 
states are being counted; for that, one needs a detailed microscopic theory. 
It does, however, fix the actual microscopic density of states: we have not 
merely reproduced black hole thermodynamics, but have made a genuine statement 
about the underlying statistical mechanics.  

Furthermore, this derivation gives us \emph{some} information about the 
microscopic states.  In standard approaches to quantum gravity, physical 
states are annihilated by all of the constraints, including the diffeomorphism 
constraint $C_\parallel^*$.  But the appearance of a central extension in the 
constraint algebra makes this impossible: the condition $C_\parallel^*
|\mathit{phys}\rangle=0$ is inconsistent with the brackets (\ref{c1a}).  One 
must instead impose a less restrictive condition, for example
\beq
\langle\mathit{phys}|C_\parallel^*|\mathit{phys}\rangle = 0 ,
\label{c6}
\eeq
which can be satisfied by states that would otherwise be discarded as 
``pure gauge.''  We thus confirm the picture of \cite{Carlip1}, that black 
hole entropy comes from ``would-be pure gauge degrees of freedom'' that 
become physical because boundary conditions relax the constraints.

These results agree with those of Ref.\ \cite{Carlip4}, although our central 
charge (\ref{b6}) and classical ``energy'' (\ref{c4}) differ by (opposite) 
factors of two from that paper.  The difference can be traced to the fact 
that Virasoro generator (\ref{b4a}) of this paper \emph{nearly} generates 
the near-horizon conformal symmetry of \cite{Carlip4}, but with a factor 
of $1/2$.  This, in turn, may reflect a difference in the choice of 
what is held fixed at the stretched horizon---in \cite{Carlip4}, $l^a$ is 
fixed---but a more detailed understanding would be desirable.  In particular, 
it has recently been shown that a near-horizon conformal symmetry exists for 
any Killing horizon \cite{Visser}; one might hope that this symmetry could 
be related to the existence of horizon constraints of the type explored 
here.  

A key question is how sensitive this derivation is to the details 
of the stretched horizon constraints.  Some flexibility certainly exists.  
One can, for instance, choose $a\ne-2$ in (\ref{b1}), and remove the resulting 
anomaly in the $\{C_\parallel,C_\pi\}^*$ bracket by defining an ``improved'' 
generator ${\widehat C}_\parallel\sim C_\parallel + bC_\pi{}^\prime$.  The 
resulting central charge then agrees with (\ref{b6}) for any value of $a$.  
Still, a more systematic understanding would be helpful.  A possible avenue 
would be to repeat this analysis in a genuine light-cone quantization at 
the true horizon, thus removing any ambiguity in the definition of the
stretched horizon. The appearance of new second class constraints makes 
such a program difficult \cite{Goldberg,Torre}, but it might be manageable 
in our relatively simple two-dimensional setting.

While the final expression (\ref{c5}) for the entropy is independent of the 
``stretching parameters'' $\epsilon_1$ and $\epsilon_2$, one might worry
that the central charge and the eigenvalue $\Delta$ depend on $\epsilon_2$,
and that the central charge becomes very small as we approach the true horizon.
This behavior may indicate that we are missing some important physics; one might
hope that an analysis in light-cone quantization, for example, will give a 
cutoff-independent value for the central charge.  It is worth noting, though,
that in a one-dimensional conformal theory with no obvious preferred periodicity,
$c$ and $\Delta$ do not have clear independent physical meanings: as Ba{\~n}ados
has observed \cite{Banados}, the transformation
\beq
L_n \rightarrow \frac{1}{k}L_{kn}
\label{c7}
\eeq
gives a new Virasoro algebra with $c\rightarrow kc$ and $\Delta\rightarrow\Delta/k$,
while preserving the density of states (\ref{c1}).  A similar ambiguity occurs
in other approaches to near-horizon state-counting, such as those of \cite{Carlip3,%
Solodukhin,Navarro,Giacomini,Carlip4}, in which $c$ and $\Delta$ each depend on an 
arbitrary parameter but have an unambiguous product.

It would also be of interest to extend this picture to the Euclidean analysis
of Ref.\ \cite{CarTeit}, to explore its relationship to the path integral approach 
to black hole entropy.  The constraint $K_2$ takes a particularly natural 
form in that setting.  The analog of the expansion is $\vartheta=n^a\nabla_aA/A$, 
while $\kappa = n^a\nabla_aN$, where $n^a$ is the unit radial normal and $N$ 
is the lapse.  Our stretched horizon constraint then describes circles in 
the $r$--$t$ plane of constant proper distance $\rho$ from the horizon, with 
$\epsilon_2 = (\partial_rA_+/A_+)\rho$.  

Finally, let me briefly address the limitations coming from the restriction to
a two-dimensional model.  While it would clearly be desirable to extend this 
analysis to higher dimensions, there are good reasons to expect that dilaton 
gravity captures the essential elements.  Indeed, general relativity in any 
dimension can be dimensionally reduced to a two-dimensional model near the
horizon of a black hole, and there are strong indications that the resulting
Liouville-like theory captures the salient features \cite{Solodukhin,Giacomini,%
Chen}.  The Euclidean path integral approach \cite{CarTeit} similarly indicates 
that the dynamics of the conjugate variables $A$ and $\kappa$ in the $r$--$t$ 
plane determines the entropy.  It is thus plausible that the two-dimensional
constraint analysis developed here will extend to black holes in any dimension.

\vspace{1.5ex}
\begin{flushleft}
\large\bf Acknowledgments
\end{flushleft}
This work was supported in part by Department of Energy grant
DE-FG03-91ER40674.

\appendix
\section{The Stretched Horizon}

In this appendix I describe the stretched horizon determined by the constraints
(\ref{b1}) in more detail.  A general two-dimensional nonextremal black hole with 
a horizon at $r=r_+$ can be described in Kruskal-like coordinates by a metric
\beq
ds^2 = -2H(r)dUdV \qquad \hbox{with} \quad UV = 2\kappa(r-r_+)J(r) ,
\label{apa}
\eeq
where $H$ and $J$ are finite and regular at the horizon.  To switch to coordinates
$(u,v)$ for which $u=0$ can be spacelike, let
\beq
U = u + \epsilon_1f(v), \quad V=v .
\label{apb}
\eeq
From (\ref{a1}) and (\ref{a4a}), it is easy to see that $\beta=0$, $\sigma\tau=H$,
and $\alpha\tau=\epsilon_1Hf'$, or with $K_1=0$,
\beq
\alpha=\epsilon_1, \quad \beta=0, \quad \sigma = 1/f', \quad \tau = Hf' .
\label{apc}
\eeq
Note that the induced metric at $u=0$ is $ds^2=-2\epsilon_1Hf'dv^2$, which will
be spacelike as long as $\epsilon_1 f'>0$, and nearly null as long as 
$\epsilon_1\ll1$.

We next compute the momentum $\pi_A$.  Equation (\ref{a6}) yields
\beq
\pi_A = \frac{2f^{\prime\prime}}{f'} + 2\epsilon_1f'\frac{\dot H}{H} \ .
\label{apd}
\eeq
Moreover, since $H$ is a function of $r$, and therefore of $UV$,
$\dot H$ and $H'$ are not independent:
\beq
{\dot H} = \frac{v}{\epsilon_1(vf)'}H'
\label{ape}
\eeq
at $u=0$.  Inserting (\ref{apd}) and (\ref{ape}) into the constraint $K_2=0$,
we obtain
\beq
\frac{A'}{\epsilon_2A_+} - \frac{f^{\prime\prime}}{f'} 
  - \frac{vf'}{(vf)'}\frac{H'}{H} = 0 .
\label{apf}
\eeq

Now assume that our $u=0$ surface is initially near the horizon, and write
\beq
2\kappa(r-r_+) = -\epsilon_2x.
\label{apg}
\eeq
Then from (\ref{apa}) and (\ref{apb}),
\beq
UV = \epsilon_1vf = 2\kappa J(r)(r-r_+) = -\epsilon_2J(r)x
\label{aph}
\eeq
at $u=0$, and
\beq
r' = -\frac{\epsilon_2}{2\kappa}x', \quad A' = -\epsilon_2\frac{\partial_rA}{2\kappa}x',
\quad H' = -\epsilon_2\frac{\partial_rH}{2\kappa}x' .
\label{api}
\eeq
The constraint (\ref{apf}) then becomes
\beq
\frac{\partial_rA}{2\kappa A_+}x' + \frac{f^{\prime\prime}}{f'}
  + \epsilon_2\frac{vf'}{(vf)'}\frac{\partial_rH}{2\kappa H}x' = 0 .
\label{apj}
\eeq
To lowest order, we see that the constraint is independent of $H$---this is a 
sort of Rindler approximation---and (\ref{apj}) is easily integrated:
\beq
f'(v) = c\exp\left\{-\frac{\partial_rA}{2\kappa A_+}x\right\}
  = c\exp\left\{\frac{\epsilon_1}{\epsilon_2}\frac{\partial_rA}{2\kappa A_+}
    \frac{1}{J_+}vf\right\} .
\label{apk}
\eeq

\begin{center}
\begin{figure}
\begin{center}
\mbox{\scalebox{.9}{%
\includegraphics*{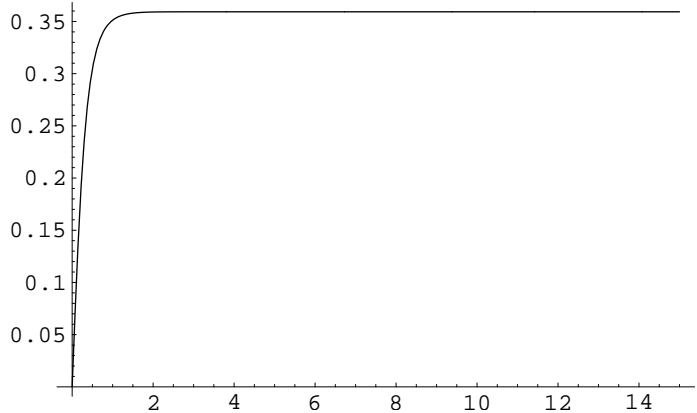}}}
\caption{$y'=e^{-axy}$ with $y(0)=0$ and $a=10$ \label{fig1}}
\end{center}
\end{figure}
\end{center}

If $\epsilon_2<0$, equation (\ref{apk}) is of the form $y'=e^{-axy}$.  Numerical 
integration shows that $f$ initially rises linearly, but very rapidly levels off 
to a nearly constant value $f_0$ of order $a^{-1/2}$ (see figure \ref{fig1}).  
If $f$ is initially near zero, its asymptotic value is
\beq
f_0\sim \left(\frac{\epsilon_1}{|\epsilon_2|}\frac{\partial_rA}{2\kappa A_+}
    \frac{1}{J_+}\right)^{-1/2} .
\label{apl}
\eeq
By (\ref{apb}), $u=0$ corresponds to $U = \epsilon_1f(V)$; our stretched horizon 
thus departs from the true horizon $U=0$, but rapidly approaches a lightlike 
surface $U=\epsilon_1f_0\propto\sqrt{\epsilon_1|\epsilon_2|}\ll1$.

\end{document}